\documentstyle[11pt,mrs2001,epsfig,rotating]{article}
\begin{document}

\def\sq{\hbox{\rlap{$\sqcap$}$\sqcup$}}
\title{THE EVOLUTION OF GALAXY CLUSTERING SINCE $z=1$}

\author{S.PHLEPS$^1$, K.MEISENHEIMER$^1$}
\affil{$^1$Max-Planck-Institut f\"ur Astronomie, K\"onigstuhl 17,
69121 Heidelberg, Germany}

\begin{abstract}
We present results of an investigation of clustering evolution 
of field galaxies between a redshift of $z\sim1$ and the present
epoch. The current analysis relies on a sample of $\sim3600$ 
galaxies  from the {\bf C}alar {\bf A}lto {\bf D}eep {\bf I}maging
{\bf S}urvey (CADIS). The multicolor classification and redshift
determination is reliable up to 
$I=23^{mag}$. The redshift distribution extends to $z\sim1.1$, the
resolution is $\Delta cz=12000$\,km~s$^{-1}$. Thus the amplitude
of the three-dimensional correlation function has to be estimated by
means of the projected correlation function $w(r_p)$. The validity
of the deprojection was tested on the Las Campanas Redshift Survey
(LCRS). The LCRS also serves as ''local'' measurement. We invented a
new method  to overcome the influence of redshift errors on
$w(r_p)$.  For evolution of the clustering
strength the ansatz $\xi(r_{com},z)\propto(1+z)^q$ is used. For the
galaxies as a whole the evolution parameter turns out to be
$q\approx-1.9$, according to the prediction of linear theory.
A formal dependency on the cosmology is presumably due to the small
number of fields observed. However, the measured clustering growth
clearly depends on Hubble type. At $z\sim1$ early type 
galaxies are already much stronger clustered, an increase with
$q\simeq-1$ is sufficient to explain the present day amplitude of the
correlation function. 
\end{abstract}

\section{The {\bf C}alar {\bf A}lto {\bf D}eep {\bf I}maging
{\bf S}urvey}

CADIS combines a very deep
emission line survey carried out with an imaging Fabry-Perot
interferometer with a deep multicolour survey using three broad-band
optical to NIR filters and up to thirteen medium-band filters when
fully completed. The combination of different observing strategies
facilitates not only the detection of emission line objects but also
the derivation of photometric spectra of all objects in the fields
without performing time consuming slit spectroscopy.\\
\noindent For the multicolor sample, a classification scheme was
developed, which is based on template spectral energy distributions
($SED$s) \cite{Wolf}. The classification algorithm basically compares the
observed colors of each object with a color library of known
objects. This color library is assembled from observed spectra by
synthetic photometry performed on an accurate representation of the
instrumental characteristics used by CADIS. Using the minimum variance
estimator, each object is assigned a
type (star -- QSO -- galaxy), a redshift (if it is not classified as
star), and an $SED$ (in terms of a number, $0\equiv$
E0, $100\equiv$ starburst). The formal errors in this process depend on
magnitude and type of the object and are of the order of
$\sigma_z=0.017$, and $\sigma_{SED}=2$, respectively.\\
\noindent The seven CADIS fields measure $\approx 1/30~\sq\deg$ each and are
located at high Galactic latitude to avoid dust absorption and reddening.
Four of these have been fully analysed so far. We
identified 3626 galaxies with $I\leq 23$ in the redshift range
$0.2\leq z\leq 1.07$.

\section{The projected correlation function}
\begin{figure}[h]
\centerline{\psfig{figure=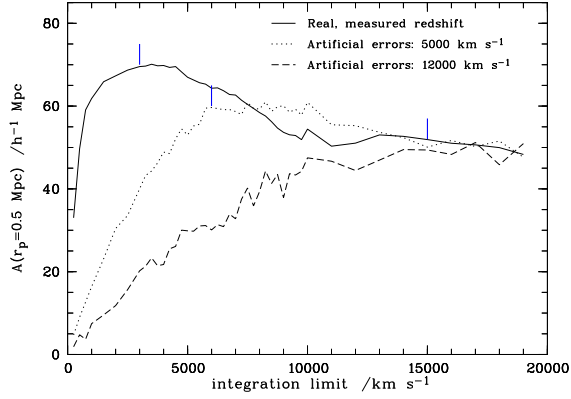,angle=270,clip=t,width=8.0cm}}
\caption[ ]{The influence of redshift measurement errors on the
projected two-point correlation function, shown is the weighted mean of  four
sectors. $w(r_p)$ for increasing integration limits is calculated for
an $\Delta cz=5000$\,km s$^{-1}$ (dotted line), and
for $\Delta cz=12000$\,km s$^{-1}$ (dashed line). The errors of the fits are not
plotted here to avoid confusion. The  marks indicate the
integration limits which have to be chosen for the calculation of
$w(r_p)$.\label{wrpmitfehlern}}
\end{figure}

\begin{figure}[h]
\unitlength1cm
\begin{turn}{270}
\begin{picture}(5.5,10)
\put(0,0){\epsfxsize=5.5cm \epsfbox{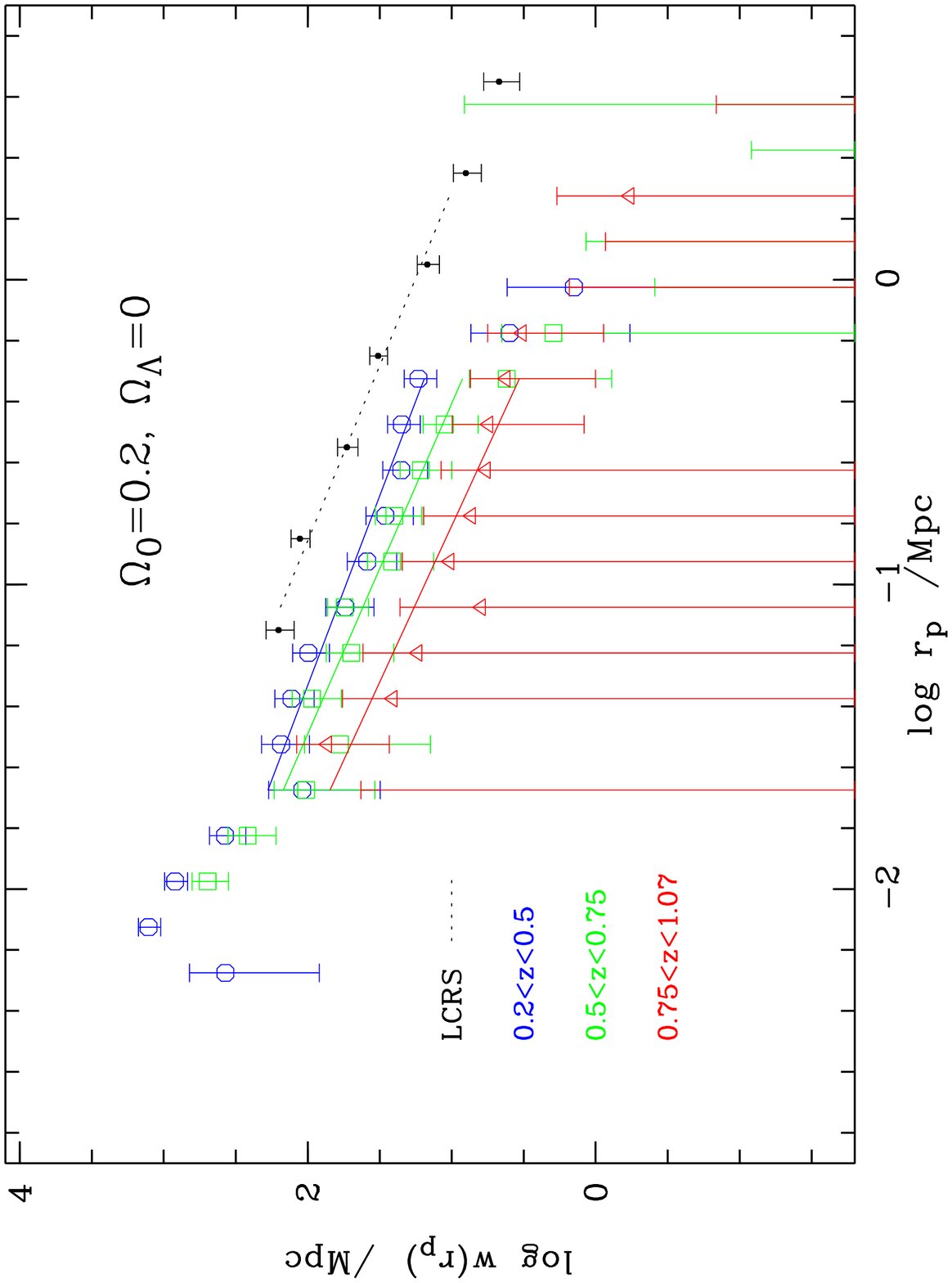}} \put(0,9){\epsfxsize=5.5cm \epsfbox{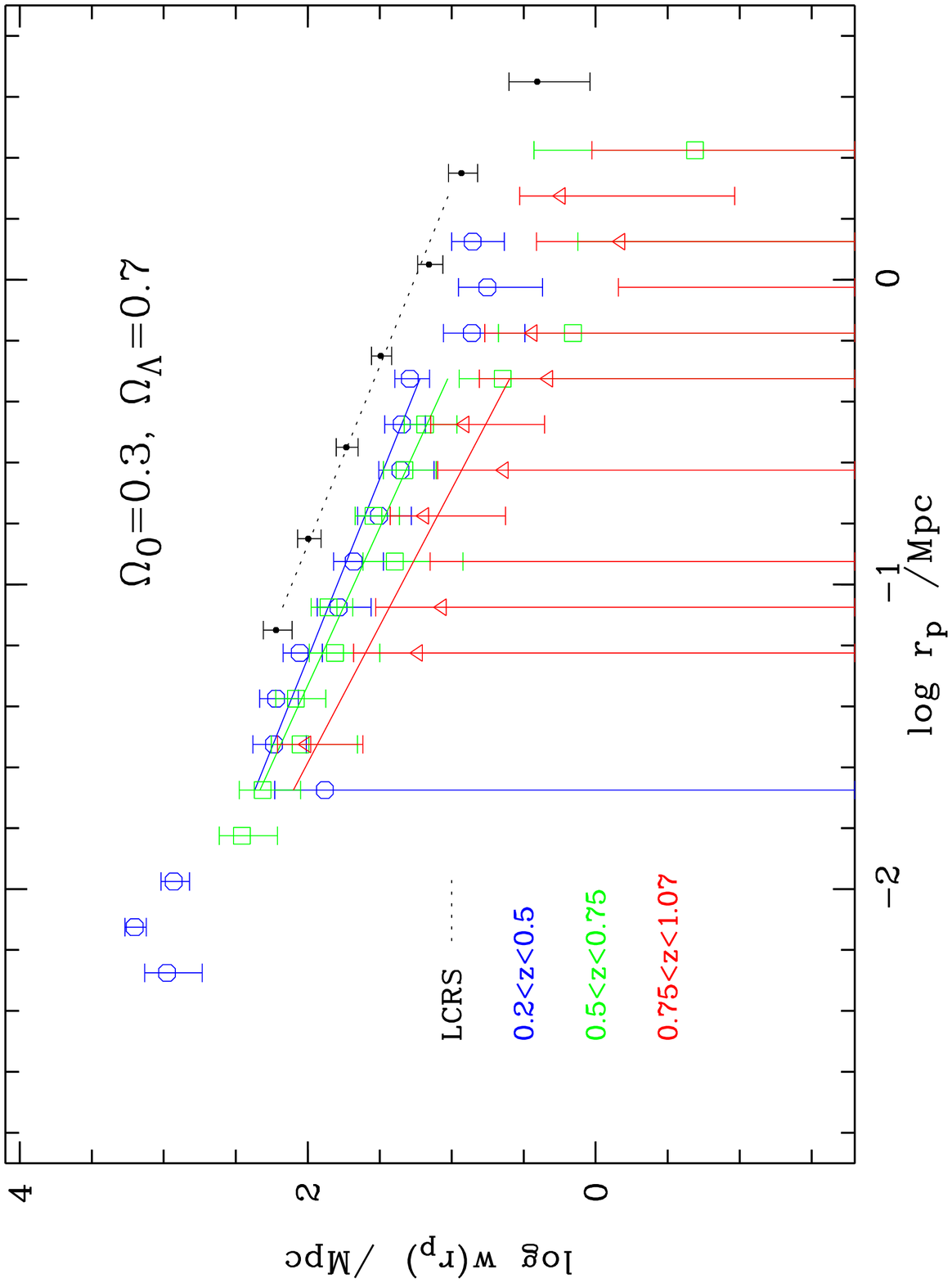}}
\end{picture}
\end{turn}
\caption{Projected correlation function in three redshift bins, for
an open model (left panel), and for a flat universe with a non-zero
cosmological constant (right panel).\label{projected}}
\end{figure}

The projected correlation function $w(r_p)$ was first introduced by
\cite{DavisPeebles} to overcome the influence of non-negligible
peculiar velocities on the three-dimensional correlation
function. Redshift inaccuracies also increase the noise and supress
the correlation signal. Therefore we used the projected correlation
function to investigate the evolution of galaxy clustering. It is defined by
\begin{eqnarray}
w(r_p)&=& 2\int_0^\infty{\xi\left[(r_p^2+\pi^2)^{1/2}\right]~{\mathrm
d}\pi}\nonumber\\
&=&2\int_{r_p}^\infty{\xi(r)(r^2-r_p^2)^{-1/2}r~{\mathrm
d}r}~.\label{projection}
\end{eqnarray}
$r_p$ is the projected distance between pairs of galaxies
(the distance perpendicular to the line of sight), $\pi$ is the distance
between the two galaxies parallel to the line of sight, and
$r^2=r_p^2+\pi^2$.
If $\xi(r)=(r/r_0)^{-\gamma}$, then equation (\ref{projection}) yields
\begin{eqnarray}\label{wrp}
w(r_p)=C r_0^\gamma r_p^{1-\gamma}
\end{eqnarray}
with
\begin{eqnarray}\label{Gamma}
C=\sqrt{\pi}\frac{\Gamma((\gamma-1)/2)}{\Gamma(\gamma/2)}~,
\end{eqnarray}
Thus computing $w(r_p)$ provides a measurement of the parameters of
the three-dimensional correlation function, namely $r_0$ and
$\gamma$.\\
\noindent Following \cite{DavisPeebles} one can calculate the
projected correlation function from
\begin{eqnarray}\label{wrpIntegration}
w(r_p)=\int_{-\delta \pi}^{+\delta \pi}{\xi(r_p,\pi)~{\mathrm
d}\pi}~.
\end{eqnarray}
Since the three-dimensional two-point correlation function has the
form of a power law, it converges rapidly to zero with increasing pair
separation. Therefore the integration limits do not have to be $\pm
\infty$, they only have to be large enough to include all correlated
pairs. Since the observable in the first place is the redshift
$z$ instead of the physical separation, we make a coordinate transform
\cite{LeFevre96}:
\begin{eqnarray}\label{wrpzInteg}
w(r_p)=\int_{-\delta z}^{+\delta z}{\xi(r_p,\pi)\frac{c}{H_0
(1+z)^2\sqrt{1+2 q_0z}}{\mathrm d}z}~,
\end{eqnarray}
for $\Omega_\Lambda=0$.\\
\noindent The way to estimate $w(r_p)$ in practice is to count the
projected distances between pairs of galaxies that are separated in
redshift space by not more than $\delta z$, in appropriate projected
distance bins. We use the estimator by \cite{LandySzalay93}, which we
will call $\zeta_{esti}(r_p)$ in the following. To derive $w(r_p)$,
$\zeta_{esti}(r_p)$ has to be multiplied with the ''effective depth''
$\Delta r_\parallel$ in which galaxies are taken into account.\\
\noindent In reality, one has to cope with a selection function of
some kind or another, and not with a top-hat function (of probability unity
to find a galaxy within the borders of the survey and zero
otherwise). The varying probability to find pairs of galaxies
separated by a redshift $\delta z$ within the survey has to be taken
into account. It can be included in the calculation by multiplying the
integrand in equation (\ref{zetaIntegration}) with
the squared redshift distribution, normalised to unity at it's
maximum. With this correction for the selection function, the
''depth'' converges to a fixed value for $\delta cz\rightarrow \infty$
and does not grow anymore, even
if the integration limits cover more than the total depth of the survey, thus
\begin{eqnarray}\label{zetaIntegration}
w(r_p)&=&\zeta_{esti}(r_p)\cdot\Delta r_\parallel\nonumber\\
&=&\zeta_{esti}\cdot\int_{-\delta z}^{+\delta
z}{\left[\frac{1}{N_{\bar{z}}}\frac{{\mathrm d}N}{{\mathrm
d}z}\right]^2\frac{c~{\mathrm d}z}{H_0(1+z)^2\sqrt{1+2q_0z}}} 
\end{eqnarray}
for $\Omega_\Lambda=0$.\\
\noindent The choice of the integration limit $\delta_z$ is somewhat
arbitrary. To find the appropriate integration limits, we simulated the
influence of redshift errors on the projected correlation function. We
added artificial redshift errors to redshifts of LCRS galaxies. The
errors (one time for a resolution of $\Delta cz=5000$\,km~s$^{-1}$
and the other time 
with $\Delta cz=12000$\,km~s$^{-1}$, the size of the CADIS redshift resolution)
were randomly drawn out of a Gaussian error distribution. We
calculated the projected correlation function for the modified samples
for increasing integration limits $\delta cz$. The amplitude was
fitted at $r_p=500 h^{-1}$\,kpc, in the range $-1.15\leq \log
r_p\leq-0.3$, to make 
sure we fit it where the signal to noise is high. Also the choice of
$r_p=500 h^{-1}$\,kpc facilitates the comparison with the CADIS data,
where we fit the amplitude at $r_p\approx316 h^{-1}$\,kpc (at the mean
redshift of the survey ($\bar{z}$) this corresponds to a
comoving separation of $\approx505h^{-1}$\,kpc).\\
\noindent Figure \ref{wrpmitfehlern} shows the amplitude of
$w(r_p=500 h^{-1}\mathrm {kpc})$, for increasing integration
limits. The measurement is compared to the modified data.\\
\noindent One starts to sample the correlation signal when the
integration limits are larger than the full width at half maximum of
the redshift error distribution. Also the amplitude is diminished. For
errors of $\Delta cz=12000$\,km~s$^{-1}$, the maximum amplitude is a factor 1.4
lower than in the case of the unchanged data.\\
\noindent We calculated $w(r_p)$ for the CADIS data with $\pm\delta cz
= 15000$\,km~s$^{-1}$. To facilitate the direct comparison with the
LCRS data, we used the modified sample (with artificial errors of size
$\Delta cz=12000$\,km~s$^{-1}$) and the same integration limits.\\
\noindent Bright stars in our field were masked out, and a random
catalogue consisting of 10000 ''galaxies'' was generated with the same
properties as the real data. The calculation was carried out for a
closed high-density model ($\Omega_0=1$, $\Omega_\Lambda=0$), a
hyperbolic low-density model 
($\Omega_0=0.2$, $\Omega_\Lambda=0$), and a flat low-density model with
non-zero comological constant ($\Omega_0=0.3$,
$\Omega_\Lambda=0.7$). Figure \ref{projected} shows the results for
the latter two ones.\\
\noindent From the amplitude of $w(r_p)$ we can deduce the amplitude of the
three-dimensional correlation function at $r_{com}=1 h^{-1}$\,Mpc:\\
\noindent With $\xi(r)=(r/r_0)^{-\gamma}$
\begin{eqnarray*}
\xi(r=1{\mathrm Mpc})&=&r_0^\gamma\\
w(r_p)&=&C r_p^{1-\gamma} r_0^\gamma\\
&=&C r_p^{1-\gamma}\xi(r=1{\mathrm Mpc})
\end{eqnarray*}
\begin{eqnarray}
\Rightarrow\xi(r_{com}=1{\mathrm Mpc})=\frac{w(r_p)(1+z)^\gamma}{C
r_p^{1-\gamma}}~,
\end{eqnarray}
with the numerical factor C from equation (\ref{Gamma}).\\
\noindent We can now parametrise the evolution with a parameter $q$
which gives directly the deviation from the global Hubble flow:
\begin{eqnarray}
\xi(r_{com}=1{\mathrm Mpc})=\xi_0(1+z)^q~.
\end{eqnarray}
$q$ can be deduced straightforwardly: if one fits $\log
\xi(r_{com}=1{\mathrm Mpc})$ versus $\log (1+z)$, it is just the
slope of the straight line.\\
\noindent We find $q=-2.68\pm0.16$ for $\Omega_0=1$, $\Omega_\Lambda=0$,
$q=-1.92\pm0.17$ for $\Omega_0=0.2$, $\Omega_\Lambda=0$, and
$q=-1.23\pm0.20$ for $\Omega_0=0.3$, $\Omega_\Lambda=0.7$. The formal
dependence on the cosmological model adopted 
for the calculation is larger than expected from a simple examination
of the distances corresponding to a certain angle $\theta$. Therefore
we regard this as an indication that the differences are presumably
due to the small number of fields observed.\\ 
\noindent We can compare our results for the $\Omega_0=1$,
$\Omega_\Lambda=0$ case with the results of \cite{LeFevre96}.
For the direct comparison we have to multiply our measured amplitudes of
the projected correlation function by 1.4 to correct for the influence
of large redshift errors.\\
\noindent With this correction, the CFRS data points are consistent with our own
measurement, although with large errors, see Figure
\ref{xiamp}. \cite{LeFevre96} claim that if 
$r_0(z=0)=5 h^{-1}$~Mpc, $0<\epsilon\la 2$. The fit of their data
points, including the connection to $z=0$, yields $q=-3.043\pm0.213$
($\cong\epsilon=1.8$). If the connection to $z=0$ is disregarded, we
find $q=-1.184\pm0.634$. The fit including our redshift error
corrected LCRS point instead of the \cite{Loveday} point yields
$q=-2.298\pm0.238$. This is even a bit less than our own measurement
($q=-2.68\pm0.16$), but nevertheless equal within the errors. 
\subsection{The evolution of clustering for different Hubble types}
\begin{figure}[h] \unitlength1cm
\begin{turn}{270}
\begin{picture}(5.5,11)
\put(0,0){\epsfxsize=5.5cm \epsfbox{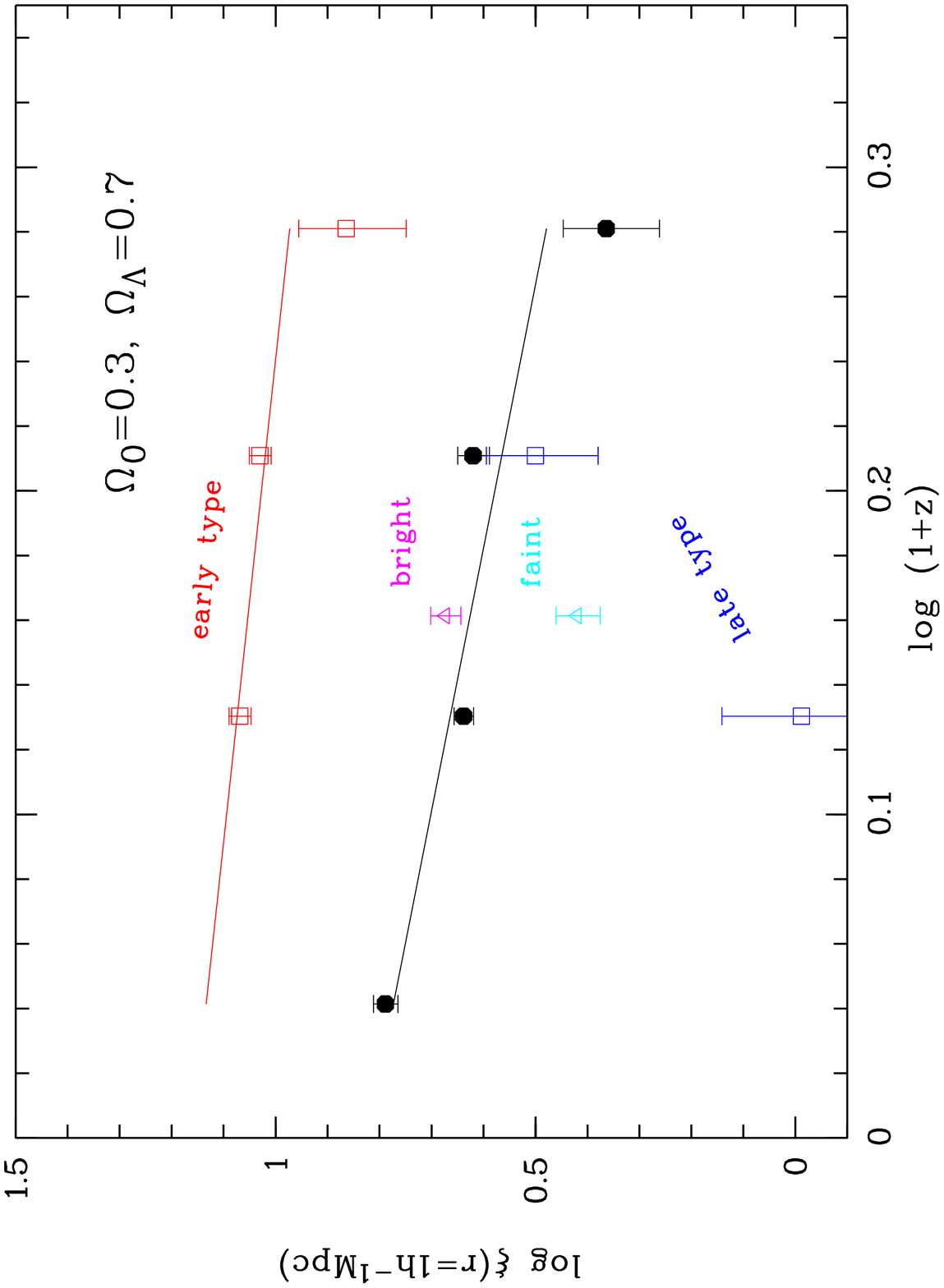}} 
\put(0,9){\epsfxsize=5.5cm \epsfbox{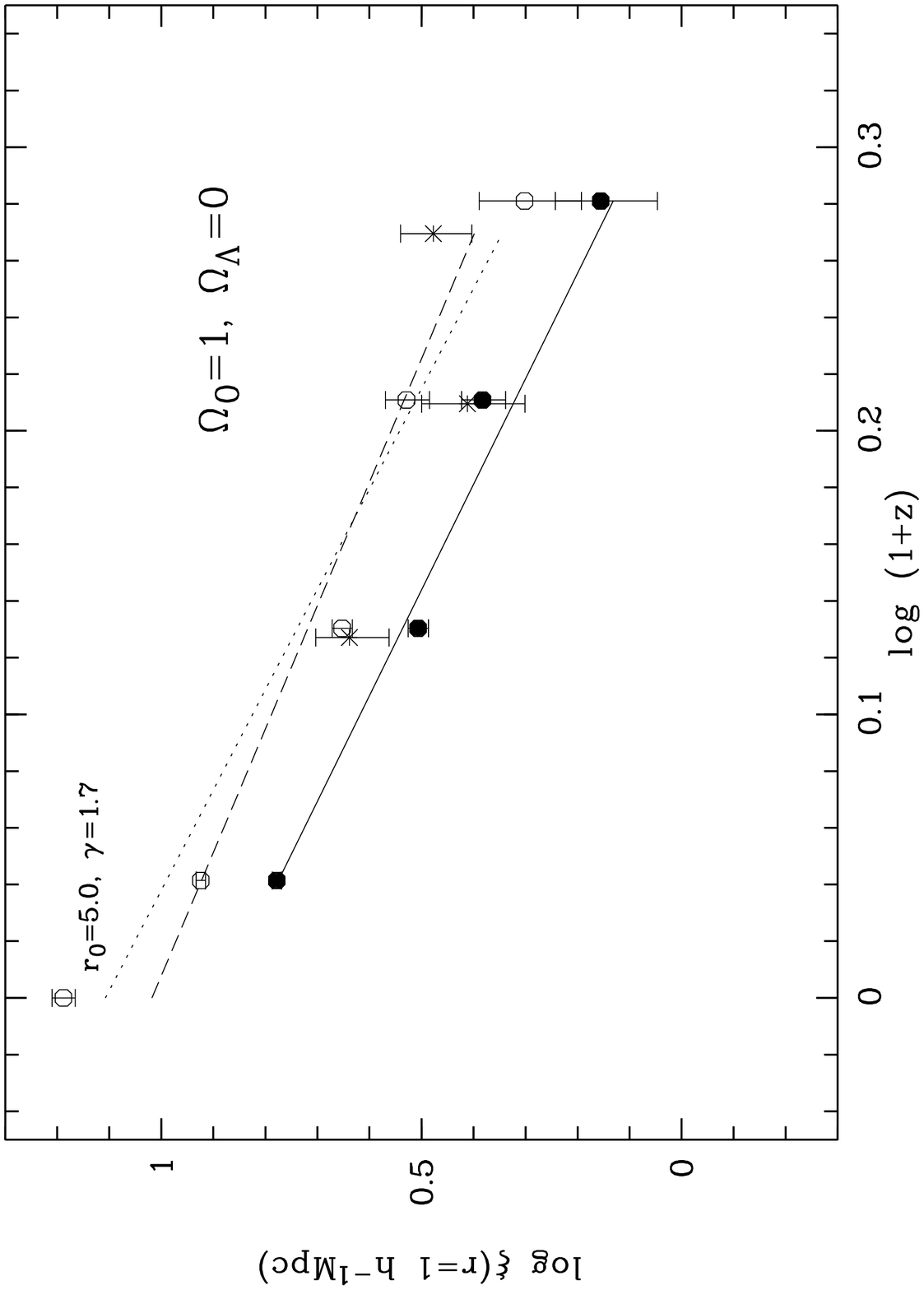}}
\end{picture}
\end{turn}
\caption{The evolution of the clustering strength (at $1 h^{-1}$\,Mpc)
with redshift. Left panel: The line is the fit to the 
data points, the first data point is the weighted mean of four LCRS
sectors, the three other ones are CADIS data, 
for a flat universe with a non-zero cosmological constant. The 
LCRS point is not included in the fit for the early type galaxies.
Right panel: The amplitudes of the three dimensional
correlation function at $r_{com}=1 h^{-1}$~Mpc, deduced from the
projected correlation function of the CFRS data (crosses), in
comparison with our own data (filled symbols). The open symbols are
our data points 
corrected for the influence of the redshift errors on the amplitude of
the projected correlation function. The dotted line is the fit of the
CFRS data points including the value of $r_0(z=0)$ from \cite{Loveday},
the dashed line is the fit using the corrected LCRS point
instead. 
\label{xiamp}}
\end{figure}

Our multicolor classification scheme gives the $SED$ of each galaxy,
and thus we are able to calculate the projected correlation function
for different Hubble types. We divided our sample at $SED=60$, the
sample with $0\leq SED\leq 60$ is called ''early type'', and the
sample with $60< SED\leq 100$ is called ''late type'' in the
following. The amplitudes of the correlation function of the early
type galaxies are at all redshifts significantly larger than for the
late type ones, and also clearly larger than for the complete CADIS
sample. Also the evolution of the clustering strength is
different. The amplitude of the correlation function of the early type
galaxies evolves much slower, we find $q=-1.71\pm0.39$ for
$\Omega_0=1$, $\Omega_\Lambda=0$, $q=-1.37\pm0.36$ for
$\Omega_0=0.2$, $\Omega_\Lambda=0$, and $q=-0.67\pm0.33$ for 
$\Omega_0=0.3$, $\Omega_\Lambda=0.7$. For the late type sample the
errors are too large to facilitate a quantitative 
statement about the custering evolution.\\
\noindent Combining $SED$ and redshift information, we are able to
calculate  restframe $B$ band luminosities for all the
galaxies. Therefore we divided our sample at $M_B=-18^{mag}$, and
calculated the projected correlation function for bright and faint
galaxies in the redshift range $0.3\leq z\leq 0.6$, where the number of faint 
galaxies is sufficient to enable the determination of the amplitude of
the projected correlation function. The bright galaxies are clearly
higher clustered than the faint ones. Figure \ref{xiamp}
shows the amplitudes of the 
three-dimensional correlation function and the fit for the evolution
paramter $q$ for the complete sample, for early and late type
galaxies and also for the bright and the faint sample. The left panel
shows the results for $\Omega_0=0.3$, $\Omega_\Lambda=0.7$, the right
panel shows the comparison of our data with the CFRS data.\\

\section{Conclusion}
We find that the amplitude of the correlation function of all the
galaxies in our sample grows with $q\simeq-1.9$ between a redshift of $z\approx
1.1$ and the present epoch. The rate of clustering growth is
consistent with the results of linear perturbation theory ($q=-2$). The
amplitude of the early type galaxies grows much 
slower with redshift ($q\simeq-1$). This result can be explained in the context of
{biased galaxy formation}, if we assume a substantial 
evolution of the galaxies between a redshift of $z\approx 1.1$ and
today. The first generation of galaxies forms in a highly clustered
state, in the bumps and wiggles superimposed on the very large scale
perturbations of the dark matter density field, whereas the next
generations of galaxies form later in the lower clustered environment 
in the wings of the super large scale overdensities. Therefore early
type galaxies are much stronger clustered than the young, late type
ones even at higher redshifts. While the universe evolves and the
clustering of the underlying dark matter density field grows, galaxies
age, and eventually merge to form massive ellipticals. They add 
to the early type population at different times, thus the early type
sample consists of galaxies of different ages, which have formed in
increasingly less clustered states. This means that biasing
decreases with increasing redshift. The net effect is a very slow
growth of clustering of early type galaxies.\\ 
\noindent Merging also plays an important role, because of two effects
on the correlation function: first of all, galaxies change their
$SED$s on relatively small timescales. They ''suddenly'' disappear from
the lower clustered late type sample and reappear in the early type
sample. The second effect is that while two galaxies merger to form
one, the small distances disappears, the probability of finding a 
galaxy near another galaxies decreases, and therefore the amplitude of
the correlation function decreases.\\
\noindent If bright galaxies form in the highest peaks of the dark
matter field, they are expected to be higher clustered than the faint
ones. The difference in clustering strength is larger for the early
type/late type sample than for bright and faint galaxies, which
corroborates the hypothesis, that galaxy evolution and merging plays
an important role in the evolution of clustering.

\vfill
\end{document}